\documentclass[showpacs,pra,onecolumn,a4paper,12pt]{revtex4}%
\usepackage[dvips]{graphics,color}
\usepackage{graphicx}
\usepackage{morefloats}
\usepackage{amsmath}
\usepackage{amsfonts}
\usepackage{amssymb}%
\providecommand{\U}[1]{\protect\rule{.1in}{.1in}}
\providecommand{\U}[1]{\protect\rule{.1in}{.1in}}
\providecommand{\U}[1]{\protect\rule{.1in}{.1in}}
\providecommand{\U}[1]{\protect\rule{.1in}{.1in}}
\providecommand{\U}[1]{\protect\rule{.1in}{.1in}}
\providecommand{\U}[1]{\protect\rule{.1in}{.1in}}
\providecommand{\U}[1]{\protect\rule{.1in}{.1in}}
\begin{document}
\title{Quasi-perfect state transfer in a bosonic dissipative network }
\author{A. Cacheffo}
\email{cacheffo@df.ufscar.br}
\author{M. A. de Ponte}
\email{maponte@df.ufscar.br}
\affiliation{Departamento de F\'{\i}sica - Universidade Federal de S\~{a}o Carlos, Caixa
Postal 676 S\~{a}o Carlos, 13565-905 SP Brazil}
\author{M. H. Y. Moussa}
\email{miled@ifsc.usp.br}
\affiliation{Instituto de F\'{\i}sica de S\~{a}o Carlos - Universidade de S\~{a}o
Paulo,Caixa Postal 369 S\~{a}o Carlos, 13560-590 SP Brazil}
\author{A. S. M. de Castro}
\email{asmcastro@uepg.br}
\affiliation{Instituto de F\'{\i}sica de S\~{a}o Carlos - Universidade de S\~{a}o
Paulo,Caixa Postal 369 S\~{a}o Carlos, 13560-590 SP; Departamento de
F\'{\i}sica, Universidade Estadual de Ponta Grossa, CEP 84030-900, Ponta
Grossa, PR, Brazil.}

\begin{abstract}
In this paper we propose a scheme for quasi-perfect state transfer in a
network of dissipative harmonic oscillators. We consider ideal sender and
receiver oscillators connected by a chain of nonideal transmitter oscillators
coupled by nearest-neighbor resonances. From the algebraic properties of the
dynamical quantities describing the evolution of the network state, we derive
a criterion, fixing the coupling strengths between all the oscillators, apart
from their natural frequencies, enabling perfect state transfer in the
particular case of ideal transmitter oscillators. Our criterion provides an
easily manipulated formula enabling perfect state transfer in the special case
where the network nonidealities are disregarded. By adjusting the common
frequency of the sender and the receiver oscillators to be out of resonance
with that of the transmitters, we demonstrate that the sender's state tunnels
to the receiver oscillator by virtually exciting the nonideal transmitter
chain. This virtual process makes negligible the decay rate associated with
the transmitter line on the expenses of delaying the time interval for the
state transfer process. Apart from our analytical results, numerical
computations are presented to illustrate our protocol.

\end{abstract}

\maketitle

\section{Introduction}

A great deal of attention has been devoted recently to the subject of perfect
state transfer (PST) in quantum networks. Since an actual quantum
processor\ would require, in fact, the ability to transfer quantum information
between spatially separated interacting systems composing a network, protocols
have been established for PST in many different and general contexts. Among
the several interesting theoretical contributions, aiming to advance the
understanding and eventual implementation of a quantum processor, the
construction of effective two-qubit gates from PST between distant nodes in
engineered bosonic and fermionic networks was proposed \cite{YungBose}. A
general formalism of the problem of PST in networks of any topology and
coupling configuration was also developed \cite{KostakNJ}. Focusing on spin
chains, PST has been pursued in networks extending beyond the nearest-neighbor
couplings \cite{Kay}, and a class of qubit networks allowing PST of any state
in a fixed period of time has been devised \cite{ChristandlDEL}. The problem
of the scaling of errors (arising from network nonidealities) with the length
of the channel connecting the nodes has been discussed \cite{Briegel}. In this
connection, a protocol for arbitrary PST in the presence of random
fluctuations in the coupling strengths of a spin chain was reported
\cite{BurgarthBose}. The entanglement dynamics in spin chains subject to noise
and disorder has also been analyzed in Ref. \cite{TsomokosHHP}. Regarding PST
in networks of harmonic oscillators, a comprehensive analysis of this subject
has been presented in Ref. \cite{PlenioHE}, and a protocol for high-efficiency
transfer of quantum entanglements in translation-invariant quantum chains has
been proposed \cite{PlenioSemiao}. In contrast to the achievements with spin
chains, there have been no proposals, until this paper, for arbitrary PST
through a network of nonideal harmonic oscillators.

Since PST is achieved by appropriate tuning of the intermode coupling and the
frequencies of the systems composing the network, the search for the general
rules governing these adjustments is of crucial interest. In this regard, it
is worth noting that the general recipe presented in Ref. \cite{KostakNJ}
contrasts with the procedure in Ref. \cite{PlenioSemiao} where a degenerate
chain of oscillators is considered: all oscillators having the same frequency
and interacting with each other with the same coupling strength. The lack of
PST in Ref. \cite{PlenioSemiao} is compensated by a more realistic possibility
of implementation of a translation-invariant network, apart from the high
efficiency transfer operation reported. We observe that in the present
contribution we also derive general conditions for PST, with the advantage
that they are easier to handle mathematically than those outlined in Ref.
\cite{KostakNJ}.

As well as in spin chains, significant advances have been made recently in
bosonic networks \cite{Audenaert,Castro,Cramer,Chow}. A general treatment of a
network of coupled dissipative quantum harmonic oscillators has been presented
recently , for an arbitrary topology, i.e., irrespective of the way the
oscillators are coupled together, the strength of their couplings, and their
natural frequencies \cite{MickelRedeGeral}. Regarding the dissipative
mechanism, two different scenarios are considered in Ref.
\cite{MickelRedeGeral}: a more realistic in which each oscillator is coupled
to its own reservoir and another with all the network oscillators coupled to a
common reservoir. Within such a general treatment of dissipation, the
emergence of relaxation- and decoherence-free subspaces in networks of weakly
and strongly coupled resonators has also been addressed \cite{MickelRDFS}. We
finally mention the proposition of a quantum memory for the preservation of
superposition states against decoherence by their evolution in appropriate
topologies of such dissipative bosonic networks \cite{MickelStoring}.

Building on the achievements mentioned above, in the present study we develop
a protocol for quasi-perfect state transfer (QPST) in a linear network of
dissipative oscillators. More specifically, we envisage the transfer of a
state between ideal sender and receiver oscillators through a linear chain of
nonideal transmitter oscillators. By analogy with Refs.
\cite{MickelRedeGeral,MickelRDFS}, here we again adopt the general and more
realistic scenario where each transmitter oscillator is coupled to a distinct
reservoir. Therefore, our goal resembles that of arbitrary PST in Ref.
\cite{BurgarthBose}; however, instead of fluctuations in the couplings of a
spin chain, we deal with the fluctuations injected by each of the reservoirs
coupled to the transmitter oscillators. Anticipating our strategy to achieve
QPST despite these sources of nonideality, we adjust the frequency of the
sender and the receiver oscillators to be significantly out of resonance with
that of the transmitters; within such an arrangement the state to be
transferred occupies the transmitter oscillators only virtually, weakening the
undesired effects of their decay mechanisms.

We point out that the recent proposal of a variety of resources for wiring up
quantum systems \cite{Wiring} lends a strongly realistic bias to the
possibility of controlling the transfer of information in quantum networks.
Following the mastering of the manipulation of the interaction between single
atoms and vibrating modes of high-Q cavities \cite{Haroche} and trapped ions
\cite{Wineland}, circuit cavity quantum electrodynamics and photonic crystals
seem to enhance the ability to transfer quantum information to a level
enabling the implementation of a logic processor \cite{LogicProcessor}. In
this connection, the elaboration of schemes to circumvent the noise injection
in the processes of PST in networks of coupled nonideal quantum systems is
indispensable, enabling protocols for fault-tolerant information transfer and
deepenning our understanding of fundamental quantum phenomena such as
entanglement and decoherence.

The plan of the paper is as follows. In Section \ref{sec2}, by analogy with
the developments presented in Ref \cite{MickelRedeGeral}, we introduce our
model in which two ideal oscillators are connected by a transmission line of
nonideal oscillators. The master equation describing the network dynamics is
presented, together with its solution. In Section \ref{sec3}, we present a
criterion for PST (QPST) in ideal (nonideal) networks of harmonic oscillators,
together with a particular application where the sets of parameters $\left\{
\omega_{m}\right\}  $ and $\left\{  \lambda_{mn}\right\}  $ ensuring PST are
derived. Our criterion relies on the definition of matrix $\mathbf{\Theta}%
(t)$, describing the evolution of the system, which serves two purposes: to
ensure PST and to compute the exchange time $t_{\mathrm{ex}}$. The analytical
treatment of QPST in small nonideal linear networks of $3$, $4$, $5$, and $6$
oscillators are given in Section \ref{sec4}. In section \ref{sec5} we apply
numerical procedures to extend our results to large numbers of nonideal
transmitters, providing a comprehensive analysis of all the network parameters
involved and demonstrating the robustness of our model. Finally, we present
our concluding remarks in Section \ref{sec6}.

\section{The model, the corresponding master equation, and its solution}

\label{sec2}

\subsection{General bosonic dissipative network}

Before introducing our model for PST, we first revisit the developments in
Ref. \cite{MickelRedeGeral}, considering a network of $N$ interacting
dissipative oscillators from the general perspective where each oscillator
interacts with each other. Any particular network topology (or graph) follows
from this general approach, with an appropriate choice of the parameters
defining the Hamiltonian modelling the network. As stressed above, by topology
is meant $i)$ which resonators are coupled together, $ii)$ their coupling
strengths and $iii)$ their natural frequencies. Moreover, in a more realistic
approach for most physical systems, it is assumed that each network oscillator
interacts with its own reservoir, instead of the special case where all the
oscillators interact with a common reservoir. Therefore, assuming from here on
that the indexes $m$, $n$, and $\ell$, labeling the oscillators, run from $1$
to $N$, we start from the general Hamiltonian
\begin{align}
\mathcal{H}=\hbar &  \sum_{m}\left[  \sum_{n}a_{m}^{\dagger}H_{mn}a_{n}%
+\sum_{k}\varpi_{mk}b_{mk}^{\dagger}b_{mk}\right. \nonumber\\
&  +\left.  \sum_{k}V_{mk}(b_{mk}^{\dagger}a_{m}+b_{mk}a_{m}^{\dagger
})\right]  , \label{1}%
\end{align}
where $b_{mk}^{\dagger}$ ($b_{mk}$) is the creation (annihilation) operator
for the $k$th bath mode $\omega_{mk}$ coupled to the $m$th network oscillator
$\omega_{m}$, whose creation (annihilation) operator is $a_{m}^{\dagger}$
($a_{m}$). The coupling strengths between the oscillators are given by the set
$\left\{  \lambda_{mn}\right\}  $, while those between the oscillators and
their reservoirs by $\left\{  V_{mk}\right\}  $. The reservoirs are modeled as
a set of $k=1,\ldots,\infty$ modes, and the elements $H_{mn}$ defining the
network topology compose the matrix
\begin{equation}
\mathbf{H}=%
\begin{pmatrix}
\omega_{1} & \lambda_{12} & \cdots & \lambda_{1N}\\
\lambda_{12} & \omega_{2} & \cdots & \lambda_{2N}\\
\vdots & \vdots & \ddots & \vdots\\
\lambda_{1N} & \lambda_{2N} & \cdots & \omega_{N}%
\end{pmatrix}
. \label{2}%
\end{equation}

To obtain the master equation of the network we first diagonalize the
Hamiltonian $\mathbf{H}$ through a canonical transformation $\ \ A_{m}%
=\sum_{n}C_{mn}a_{n}$, where the coefficients of the $m$th line of matrix
$\mathbf{C}$ define the eigenvectors associated with the eigenvalues
$\varpi_{m}$ of matrix (\ref{2}). The commutation relations $\left[
A_{m},A_{n}^{\dag}\right]  =\delta_{mn}$ and $\left[  A_{m},A_{n}\right]  =0$,
following from the orthogonality of matrix $\mathbf{C}$, in that
$\mathbf{C}^{T}=\mathbf{C}^{-1}$, enable us to rewrite Hamiltonian (\ref{1})
in terms of decoupled normal-mode oscillators $\varpi_{m}$, each of them
interacting, however, with all the $N$ reservoirs. The diagonalized
$\mathbf{H}$ helps us to introduce the interaction picture, within which a set
of assumptions leads to the master equation describing the network evolution.
We first assume that the couplings between the normal-mode oscillators and the
reservoirs are weak enough to allow a second-order perturbation approximation.
Moreover, under Markovian white noise, the evolved reduced density operator of
the network, $\rho(t)$, is assumed to be factorized from the stationary
density operator of the reservoirs. Finally, the reservoir frequencies are
assumed to be closely spaced enough to allow a continuum summation, the
spectral density $\sigma_{m}(\varpi_{n})$ and coupling parameter $V_{m}%
(\varpi_{n})$ being slowly varying functions. Thus, after tracing out the
degrees of freedom of the absolute zero reservoirs, we obtain the generalized
Lindblad form%
\begin{align}
\frac{\operatorname*{d}\rho(t)}{\operatorname*{d}t}  &  =\sum_{m,n}\left\{
\frac{i}{\hbar}\left[  \rho(t),a_{m}^{\dagger}H_{mn}a_{n}\right]  \right.
\nonumber\\
&  +\left.  \frac{\Gamma_{mn}}{2}\left(  \left[  a_{n}\rho(t),a_{m}^{\dag
}\right]  +\left[  a_{m},\rho(t)a_{n}^{\dag}\right]  \right)  \right\}
\nonumber\\
&  \equiv\sum_{m,n}\left\{  \frac{i}{\hbar}\left[  \rho(t),a_{m}^{\dagger
}H_{mn}a_{n}\right]  +\mathcal{L}_{mn}\rho(t)\right\}  \mathrm{,} \label{3}%
\end{align}
where we have defined the effective damping matrix $\mathbf{\Gamma}$, whose
elements are%
\[
\Gamma_{mn}=N\sum_{n^{\prime}}C_{n^{\prime}m}\gamma_{m}(\varpi_{n^{\prime}%
})C_{n^{\prime}n}\text{,}%
\]
with $\gamma_{m}(\varpi_{n})=\frac{1}{N}\left[  V_{m}(\varpi_{n})\sigma
_{m}(\varpi_{n})\right]  ^{2}\int_{-\varpi_{n}}^{\infty}\delta\left(
\epsilon\right)  \operatorname*{d}\epsilon$. In Eq. (\ref{3}),$\ $the
Liouville operators $\mathcal{L}_{mn}\rho(t)$ account for both the direct
($m=n$) and indirect ($m\neq n$)\ dissipative channels. Through the direct
dissipative channels, the oscillators lose excitation to their own reservoirs,
at a damping rate $\Gamma_{mm}$, whereas through the indirect channels they
lose excitation to all the other reservoirs but not to their own. For
Markovian white noise reservoirs the indirect channels disappear, since the
spectral densities of the reservoirs are invariant over translation in
frequency space, rendering $\gamma_{m}(\varpi_{n^{\prime}})=\gamma_{m}$, and,
consequently, $\Gamma_{mn}=N\gamma_{m}\delta_{mn}$ \cite{Mickel1,Mickel2}.

It is worth stressing that the whole of the working presented in the next
subsection, where a solution to the master equation is derived, applies to the
case of non-Markovian reservoirs, thus including the indirect dissipative
channels. Markovian reservoirs are only assumed at the end of subsection C,
when the topology of our framework is finally defined as a linear chain of
dissipative oscillators.

\subsection{Solution of the master equation}

To obtain a solution of Eq. (\ref{3}), we shall employ it to derive the
Glauber-Sudarshan $P$ function for the network%
\begin{equation}
\frac{\operatorname*{d}P(\{\eta_{m^{\prime}}\},t)}{\operatorname*{d}t}%
=\sum_{m}\left(  \frac{\Gamma_{mm}}{2}+\sum_{n}H_{mn}^{D}\eta_{n}%
\frac{\partial}{\partial\eta_{m}}+c.c.\right)  P(\{\eta_{m^{\prime}%
}\},t)\text{,} \label{P}%
\end{equation}
where we have defined the matrix elements $H_{mn}^{D}=iH_{mn}+\Gamma_{mn}/2$,
extending the former $H_{mn}$, to take into account the dissipative ($D$)
process. Assuming that the general network described by matrix (\ref{2}) is
composed entirely of dissipative oscillators, the matrix $\mathbf{H}^{D}$
assumes the form%
\begin{equation}
\mathbf{H}^{D}=i\mathbf{H}+\frac{1}{2}%
\begin{pmatrix}
\Gamma_{11} & \Gamma_{12} & \cdots & \Gamma_{1N}\\
\Gamma_{21} & \Gamma_{22} & \cdots & \Gamma_{2N}\\
\vdots & \vdots & \ddots & \vdots\\
\Gamma_{N1} & \Gamma_{N2} & \cdots & \Gamma_{NN}%
\end{pmatrix}
\text{.} \label{2D}%
\end{equation}

For the initial state of the network we consider the general pure
superpositions of coherent states%
\begin{equation}
\rho(0)=\mathcal{N}^{2}\sum_{r,s=1}^{Q}\Lambda_{r}\Lambda_{s}^{\ast}\left\vert
\left\{  \beta_{m}^{r}\right\}  \right\rangle \left\langle \left\{  \beta
_{m}^{s}\right\}  \right\vert \text{,} \label{Rho0}%
\end{equation}
where $\mathcal{N}$ is the normalization factor, $\Lambda_{r}$ is the
probability amplitude of the product state $\left\vert \left\{  \beta_{m}%
^{r}\right\}  \right\rangle =\bigotimes\nolimits_{m=1}^{N}\left\vert \beta
_{m}^{r}\right\rangle $, and the labels $r$ and $s$ run from $1$\ to the
integer $Q$. The superscript $r$ stands for the $r$th state of the
superposition while the subscript $m$ stands for the coherent state of $\ $the
$m$th oscillator. (We stress that the discrete sum of product states in Eq.
(\ref{Rho0}) can be substituted by the continuum sum $\left\vert
\psi(0)\right\rangle =\mathcal{N}\int d\theta\Lambda\left(  \theta\right)
\left\vert \left\{  \beta_{m}\left(  \theta\right)  \right\}  \right\rangle $
with no further complication). From Eqs. (\ref{P}) and (\ref{Rho0}), it is
straightforward to show that the network density operator evolves as%
\begin{equation}
\rho(t)=\mathcal{N}^{2}\sum_{r,s}\Lambda_{r}\Lambda_{s}^{\ast}\frac
{\left\langle \left\{  \beta_{m}^{s}\right\}  \right.  \left\vert \left\{
\beta_{m}^{r}\right\}  \right\rangle }{\left\langle \left\{  \zeta_{m}%
^{s}(t)\right\}  \left\vert \left\{  \zeta_{m}^{r}(t)\right\}  \right.
\right\rangle }\left\vert \left\{  \zeta_{m}^{r}(t)\right\}  \right\rangle
\left\langle \left\{  \zeta_{m}^{s}(t)\right\}  \right\vert \text{.}
\label{RhoT}%
\end{equation}
The excitation of the $m$th oscillator, given by
\begin{equation}
\zeta_{m}^{r}\left(  t\right)  =\sum_{n}\Theta_{mn}(t)\beta_{n}^{r}\text{.}
\label{4}%
\end{equation}
follows from the time-dependent matrix elements
\[
\Theta_{mn}(t)=\sum_{m^{\prime}}D_{mm^{\prime}}\exp\left(  -\mathfrak{W}%
_{m^{\prime}}t\right)  D_{m^{\prime}n}^{-1}\text{.}%
\]
where the $m$th column of matrix $\mathbf{D}$ defines the $m$th eigenvector
associated with the eigenvalue $\mathfrak{W}_{m}$ of matrix $\mathbf{H}^{D}$.

For the reduced density operator of the $m$th oscillator we obtain%
\begin{equation}
\rho_{m}(t)=\mathcal{N}^{2}\sum_{r,s}\Lambda_{r}\Lambda_{s}^{\ast}%
\frac{\left\langle \left\{  \beta_{n}^{s}\right\}  \right.  \left\vert
\left\{  \beta_{n}^{r}\right\}  \right\rangle }{\left\langle \zeta_{m}%
^{s}(t)\left\vert \zeta_{m}^{r}(t)\right.  \right\rangle }\left\vert \zeta
_{m}^{r}(t)\right\rangle \left\langle \zeta_{m}^{s}(t)\right\vert \text{,}
\label{rho}%
\end{equation}
where the influence of all the other oscillators of the network is present
explicitly in the product $\left\langle \left\{  \beta_{n}^{s}\right\}
\right.  \left\vert \left\{  \beta_{n}^{r}\right\}  \right\rangle $ and
implicitly in the states $\left\vert \zeta_{m}^{r}(t)\right\rangle $.

\subsection{A linear dissipative network: our framework}

A linear ($lin$) dissipative network of nearest-neighbor interacting harmonic
oscillator is built up by coupling the $k$th oscillator with the $\left(
k\pm1\right)  $th oscillators, leaving the first oscillator ($m=1$) uncoupled
from the\ last one ($m=N$). The matrix $\mathbf{H}_{lin}^{D}$ obtained for
this case has the three-diagonal form%
\[
\mathbf{H}_{lin}^{D}=i\left(
\begin{array}
[c]{ccccccc}%
\omega_{1} & \lambda_{12} & 0 & \cdots & 0 & 0 & 0\\
\lambda_{12} & \omega_{2} & \lambda_{23} & \cdots & 0 & 0 & 0\\
0 & \lambda_{23} & \omega_{3} & \ddots & 0 & 0 & 0\\
\vdots & \vdots & \ddots & \ddots & \ddots & \vdots & \vdots\\
0 & 0 & 0 & \ddots & \omega_{N-2} & \lambda_{N-2,N-1} & 0\\
0 & 0 & 0 & \cdots & \lambda_{N-2,N-1} & \omega_{N-1} & \lambda_{N-1,N}\\
0 & 0 & 0 & \cdots & 0 & \lambda_{N-1,N} & \omega_{N}%
\end{array}
\right)  +\frac{1}{2}%
\begin{pmatrix}
\Gamma_{11} & \Gamma_{12} & \cdots & \Gamma_{1N}\\
\Gamma_{21} & \Gamma_{22} & \cdots & \Gamma_{2N}\\
\vdots & \vdots & \ddots & \vdots\\
\Gamma_{N1} & \Gamma_{N2} & \cdots & \Gamma_{NN}%
\end{pmatrix}
\text{.}%
\]

As already made clear, we focus on the case of ideal sender and receiver
oscillators, here assumed to be the first and the last, respectively, both
with the same frequency $\omega$. All the transmitter oscillators, from the
second to the ($N-1$)th, are assumed from here on to decay at the same rate
$\Gamma_{mn}=\Gamma$, and to be tuned out of resonance with the sender and
receiver, to frequency $\Omega$. Regarding the coupling between the
oscillators, we assume that the sender and the receiver are connected with
their transmitter neighbors with the same strength $\lambda$, whereas the
transmitters are connected to each other with strength $\varepsilon\lambda$,
$\varepsilon$ being a dimensionless parameter allowing the couplings within
the transmitting channel to be controlled. Finally, assuming Markovian white
noise reservoirs, to eliminate the indirect dissipative channels, the above
matrix simplifies to
\begin{equation}
\mathbf{H}_{lin}^{D}=i%
\begin{pmatrix}
\omega & \lambda & 0 & \cdots & 0 & 0 & 0\\
\lambda & \Omega-i\Gamma/2 & \varepsilon\lambda & \cdots & 0 & 0 & 0\\
0 & \varepsilon\lambda & \Omega-i\Gamma/2 & \ddots & 0 & 0 & 0\\
\vdots & \vdots & \ddots & \ddots & \ddots & \vdots & \vdots\\
0 & 0 & 0 & \ddots & \Omega-i\Gamma/2 & \varepsilon\lambda & 0\\
0 & 0 & 0 & \cdots & \varepsilon\lambda & \Omega-i\Gamma/2 & \lambda\\
0 & 0 & 0 & \cdots & 0 & \lambda & \omega
\end{pmatrix}
\text{.} \label{Lin}%
\end{equation}

\section{A criterion for PST (QPST) in ideal (nonideal) networks of harmonic
oscillators}

\label{sec3}

\subsection{PST in ideal networks}

On the basis of the model developed above, in this section we derive a general
criterion for PST, whatever the topology of an ideal ($\Gamma=0$) network of
harmonic oscillators. Such a criterion for PST is further applied for the
particular case of the linear topology modeled by Eq. (\ref{Lin}). Starting
from Eq. (\ref{4}) written as the matrix product $\mathbf{\zeta}^{r}\left(
t\right)  =\mathbf{\Theta}(t)\cdot\mathbf{\beta}^{r}$, it is straightforward
to conclude that the condition for transferring the state of the first
oscillator to the $N$th is given by the matrix structure%
\begin{equation}
\mathbf{\Theta(}t_{\mathrm{ex}}\mathbf{)=}%
\begin{pmatrix}
0 & 0\hspace{0.2in}\cdots & 1\\%
\genfrac{}{}{0pt}{0}{\genfrac{}{}{0pt}{0}{{}}{0}}{\genfrac{}{}{0pt}{0}{\vdots
}{{}}}%
& \left[  \hspace{0.2in}\underset{\underset{\underset{}{}}{}}{{\Huge \forall}%
}\hspace{0.2in}\right]  &
\genfrac{}{}{0pt}{0}{\genfrac{}{}{0pt}{0}{{}}{0}}{\genfrac{}{}{0pt}{0}{\vdots
}{{}}}%
\\
1 & 0\hspace{0.2in}\cdots & 0
\end{pmatrix}
\text{, } \label{Theta}%
\end{equation}
which reversely ensures that the state of the $N$th oscillator can equally be
transferred to the first one, no matter what happens to the states of the
intervening $N-2$ transmitting oscillators, from the second to the last but
one. That is why the submatrix formed by removing the first and last rows and
columns of $\mathbf{\Theta}$ is left undefined; for any submatrix
$[\mathbf{\forall]}$, PST is achieved as long as the states of the first and
the last oscillators are interchanged. It is worth noting that the criterion
for state exchange fixed by the structure of the matrix in Eq. (\ref{Theta})
does not depend on the network state. Evidently, the imposition of this matrix
form can be used to compute the exchange time $t_{\mathrm{ex}}$.

Since the evolution of matrix $\mathbf{\Theta}(t)$, given by
\begin{align}
\mathbf{\Theta}(t)  &  =\mathbf{D\cdot}\exp[-i\mathbf{R}t]\cdot\mathbf{D}%
^{-1}\nonumber\\
&  =\exp[-(\mathbf{D\cdot}i\mathbf{R}\cdot\mathbf{D}^{-1})t]=\exp
[-i\mathbf{H}t]\text{,} \label{Theta1}%
\end{align}
is governed by the Hamiltonian $\mathbf{H}$ (whose eigenvalues $R_{m\text{ }}%
$compose the diagonal matrix $\mathbf{R}$), the sets of parameters $\left\{
\omega_{m}\right\}  $ and $\left\{  \lambda_{mn}\right\}  $ complying with a
necessary condition for PST in a particular network topology follow from the
commutation relation
\begin{equation}
\left[  \Theta(t_{\mathrm{ex}}),\mathbf{H}\right]  =0\text{,} \label{cm}%
\end{equation}
which holds at any time, including the state exchange time $t_{\mathrm{ex}}$.
We observe that not all sets $\left\{  \omega_{m}\right\}  $ and $\left\{
\lambda_{mn}\right\}  $ derived from condition (\ref{cm}) ensure PST.
For\textbf{ }a necessary and sufficient condition we must choose, among the
sets of parameters $\left\{  \omega_{m}\right\}  $ and $\left\{  \lambda
_{mn}\right\}  $, those ensuring the reduction of matrix (\ref{Theta1}) to
(\ref{Theta}), our initial premise, at the exchange time. Hence, matrix
$\mathbf{\Theta}(t)$ can be used for two purposes: to ensure PST and to
compute the exchange time $t_{\mathrm{ex}}$.

We observed in the Introduction that our general condition for PST, given by
Eq. (\ref{cm}), has the advantage of being easier to handle mathematically
than that outlined in Ref. \cite{KostakNJ}. In fact, in Ref. \cite{KostakNJ},
the condition for PST follows from the diagonalization of a permutation matrix
---equivalent to the above defined $\mathbf{\Theta}(t)$--- whereas our
condition reduces to the computation of a commutation relation.

It must be stressed that in the case where the state transfer is achieved by
tunneling, the set of coupling parameters $\left\{  \lambda_{mn}\right\}  $
does not play a significant role in the process since the transferred state
does not effectively occupy any of the transmitter oscillators. Actually, for
the state to be transferred by tunneling, we only have to ensure that the
common frequency ($\omega$) of the sender and the receiver oscillators is
distinct from those of the transmitter ones; under this condition, we can
verify the relation (\ref{cm}) independently of the set of coupling strengths
$\lambda_{mn}$, and these may all have the same value $\lambda$.

\subsection{QPST in nonideal networks}

The extension of the above criterion to state transfer in nonideal networks,
the case at hand, follows directly by assuming a finite value of $\Gamma$. The
damping rates due to dissipation are thus introduced into Eqs. (\ref{Theta})
and (\ref{Theta1}), generalizing the relation (\ref{cm}) to
\begin{equation}
\left[  \Theta(t_{\mathrm{ex}}),\mathbf{H}^{D}\right]  =0\text{.} \label{CR}%
\end{equation}
However, in this general nonideal case the transferred state exhibits a
fidelity that decreases with time and the magnitude of the decay rate $\Gamma
$, apart from depending on other properties such as the network topology.
Moreover, there are two further situations, not envisaged in this paper, where
the commutation relation (\ref{CR}) reduces to that of the ideal case
(\ref{cm}), even when the fidelity decreases with time and the magnitude of
$\Gamma$: $i)$ when $\Gamma_{mn}=\Gamma\delta_{mn}$ and $ii)$ $\Gamma
_{mn}=\Gamma$. In both cases the sets of parameters $\left\{  \omega
_{m}\right\}  $ and $\left\{  \lambda_{mn}\right\}  $ ensuring QPST are
exactly those ensuring PST.

It is worth noting again that, confining ourselves to a linear dissipative
network, the adopted strategy of considering the ideal sender and receiver
oscillators significantly out of resonance with the nonideal transmitters,
ensures QPST despite the nonidealities of the transmitter line. In fact, as
anticipated in the Introduction and demonstrated below, the virtual occupation
of the transmitter line protects the transferred state almost perfectly from
the damping mechanisms.

\subsection{An application of the commutation relation $\left[  \Theta
(t_{\mathrm{ex}}),\mathbf{H}\right]  =0$}

Considering the Hamiltonian $\mathbf{H}$ in Eq. (\ref{2}) and the particular
choice
\begin{equation}
\Theta(t_{\mathrm{ex}})=\left(
\begin{array}
[c]{ccccc}%
0 & 0 & \cdots & 0 & 1\\
0 & 0 & \cdots & 1 & 0\\
\vdots & \vdots & \ddots & \vdots & \vdots\\
0 & 1 & \cdots & 0 & 0\\
1 & 0 & \cdots & 0 & 0
\end{array}
\right)  \label{theta}%
\end{equation}
ensuring the transfer of the initial state of the $m$th oscillator to the
$\left[  N-\left(  m-1\right)  \right]  $th one, we obtain from relation
(\ref{cm}) the condition for PST:%
\begin{align*}
\omega_{m}  &  =\omega_{N-\left(  m-1\right)  }\text{,}\\
\lambda_{mn}  &  =\lambda_{N-\left(  m-1\right)  ,N-\left(  n-1\right)
}\text{,}%
\end{align*}
which generalizes the one introduced in Ref. \cite{PlenioHE}, given by
\begin{subequations}
\label{PP}%
\begin{align}
\omega_{m}  &  =\omega\text{,}\label{Pa}\\
\lambda_{m,m+1}  &  =\lambda_{m+1,m}=\lambda\sqrt{m\left(  N-m\right)
}\text{.} \label{Pb}%
\end{align}
Evidently, each choice of the submatrix $[\mathbf{\forall]}$ in the general
form (\ref{Theta}) prompts a different set of parameters $\left\{  \omega
_{m}\right\}  $ and $\left\{  \lambda_{mn}\right\}  $ ensuring PST.

\subsection{An alternative way to compute $t_{\mathrm{ex}}$}

Another way to compute the exchange time $t_{\mathrm{ex}}$, instead of using
matrix $\Theta(t_{\mathrm{ex}})$, follows from the time profile of the
probability of a successful transfer of the desired state ---or equivalently,
the fidelity of the transfer process--- given by
\end{subequations}
\begin{equation}
\mathcal{P}(t)=\operatorname*{Tr}\left[  \rho(0)\rho(t)\right]  \text{.}
\label{F}%
\end{equation}
Evidently, in a general situation where PST is sought --- other than the
transfer by tunneling analyzed here --- the computation of the exchange time
$t_{\mathrm{ex}}$ from Eq. (\ref{F}), must take into account the set of
parameters $\left\{  \omega_{m}\right\}  $ and $\left\{  \lambda_{mn}\right\}
$ ensuring PST, derived from the commutation relation (\ref{CR}). A further
condition is that the ideal case $\Gamma=0$ must render $\mathcal{P}%
(t_{\mathrm{ex}})=1$. Although a rather demanding task, the maximization of
the probability $\mathcal{P}(t_{\mathrm{ex}})$ yields, by itself, a necessary
and sufficient condition for the derivation of the sets of parameters
$\left\{  \omega_{m}\right\}  $ and $\left\{  \lambda_{mn}\right\}  $ ensuring
PST. The commutation relation (\ref{CR}) provides a shortcut for this task.

We assume, as usual, that the initial density operator $\rho(0)$ factorizes as
$\rho_{1}(0)%
{\textstyle\bigotimes_{m=2}^{N}}
\rho_{m}(0)$, $\rho_{1}(0)$ representing the state of the first oscillator, to
be transferred to the $N$th one, and $\rho_{m}(0)$ standing for the initial
state of the transmitter and the receiver oscillators. In its turn, the
density operator at the exchange time $\rho(t_{\mathrm{ex}})$ factorizes as
$\rho_{N}(t_{\mathrm{ex}})%
{\textstyle\bigotimes_{m=1}^{N-1}}
\rho_{m}(t_{\mathrm{ex}})$, where $\rho_{N}(t_{\mathrm{ex}})$ represents the
transferred state of the $N$th oscillator, and $\rho_{m}(t_{\mathrm{ex}})$ the
final state of the transmitter oscillators plus the sender. Evidently, the
state of the $N$th oscillator at $t_{\mathrm{ex}}$ must be, in the ideal case
where $\Gamma=0$, exactly that prepared in the first oscillator. With this
assumption, by substituting Eq. (\ref{rho}) into $\mathcal{P}(t_{\mathrm{ex}%
})$, we obtain%

\begin{align}
\mathcal{P}(t_{\mathrm{ex}})  &  =\mathcal{N}^{4}\sum_{r,s,r^{\prime
},s^{\prime}}\Lambda_{r}\Lambda_{s}^{\ast}\Lambda_{r^{\prime}}\Lambda
_{s^{\prime}}^{\ast}\left\langle \left\{  \beta_{n}^{s}\right\}  \right.
\left\vert \left\{  \beta_{n}^{r}\right\}  \right\rangle \left\langle \left\{
\beta_{n}^{s^{\prime}}\right\}  \right.  \left\vert \left\{  \beta
_{n}^{r^{\prime}}\right\}  \right\rangle \nonumber\\
&  \times\exp\left\{  -\left[  \zeta_{N}^{s^{\prime}}(t_{\mathrm{ex}}%
)-\beta_{1}^{s}\right]  ^{\ast}\left[  \zeta_{N}^{r^{\prime}}(t_{\mathrm{ex}%
})-\beta_{1}^{r}\right]  \right\}  \text{.} \label{Prob}%
\end{align}

As an illustrative application of Eq. (\ref{Prob}), we consider the specific
case where the state to be transferred to the $N$th oscillator is prepared in
the first one as the Schr\"{o}dinger cat-like superposition $\mathcal{N}%
\left(  \left\vert \alpha\right\rangle _{1}+\left\vert -\alpha\right\rangle
_{1}\right)  $, while all other oscillators are in the vacuum state. We obtain
from Eq. (\ref{Prob}), for the ideal case, the relations
\begin{subequations}
\label{T}%
\begin{align}%
{\textstyle\sum\limits_{m}}
C_{Nm}\cos\left(  R_{m}t_{\mathrm{ex}}\right)  C_{m1}^{-1}  &  =\pm
1,\label{Ta}\\%
{\textstyle\sum\limits_{m}}
C_{Nm}\sin\left(  R_{m}t_{\mathrm{ex}}\right)  C_{m1}^{-1}  &  =0, \label{Tb}%
\end{align}
which enable us to determine $t_{\mathrm{ex}}$ as well as Eq. (\ref{cm}),
remembering that $R_{m\text{ }}$ stands for the eigenvalues of the free
Hamiltonian $\mathbf{H}$. Therefore, the exchange time $t_{\mathrm{ex}}$
follows from the computation of the eigenstates and the associated
eigenvectors of $\mathbf{H}$.

\section{Analytical treatment of QPST in small nonideal linear networks}

Now, focusing on our scheme for the transfer of states by tunneling , we
present an analytical treatment of small nonideal linear networks, using the
particular case of (\ref{theta}) to compute the exchange time $t_{\mathrm{ex}%
}$. As an illustration, consider the case of QPST in a network with $N=4$,
whose dissipative matrix, following from Eq. (\ref{Lin}), is%
\end{subequations}
\[
\mathbf{H}_{lin(N=4)}^{D}=%
\begin{pmatrix}
i\omega & i\lambda & 0 & 0\\
i\lambda & i\Omega+\Gamma/2 & i\varepsilon\lambda & 0\\
0 & i\varepsilon\lambda & i\Omega+\Gamma/2 & i\lambda\\
0 & 0 & i\lambda & i\omega
\end{pmatrix}
\text{.}%
\]
The eigenvalues and eigenvectors of $\mathbf{H}_{lin(N=4)}^{D}$ defines the
matrix $\mathbf{\Theta}(t)$ and, consequently, the evolved excitation of the
oscillators $\mathbf{\zeta}^{r}\left(  t\right)  =\mathbf{\Theta}%
(t)\cdot\mathbf{\beta}^{r}$. Next, we introduce the scaled damping rate
$\eta=\Gamma/\lambda$, frequency $\varpi=\omega/\lambda$, and time
$\tau=\lambda t$, apart from the detunings $\Delta_{\pm}=\Omega\pm\omega$ and
the effective coupling $\mu=\lambda/\Delta_{-}$, to stress that we must focus
on the regime where $\mu\ll1$ and $\eta\ll1$. These parameters ensure the weak
coupling between the transmitter oscillators and their respective reservoirs
---justifying the master equation derived above --- apart from enabling the
expansion of matrix $\mathbf{\Theta}(t)$ to second order in $\mu$, giving
\begin{equation}
\mathbf{\Theta}(\tau)=e^{-i(\varpi-\mu)\tau}e^{-\eta\mu^{2}\tau}%
\begin{pmatrix}
\cos(\varepsilon\mu^{2}\tau) & -\mathfrak{g}_{c}(\tau) & i\mathfrak{g}%
_{s}(\tau) & -i\sin(\varepsilon\mu^{2}\tau)\\
-\mathfrak{g}_{c}(\tau) & \mathfrak{h}_{c}(\tau) & -i\mathfrak{h}_{s}(\tau) &
i\mathfrak{g}_{s}(\tau)\\
i\mathfrak{g}_{s}(\tau) & -i\mathfrak{h}_{s}(\tau) & \mathfrak{h}_{c}(\tau) &
-\mathfrak{g}_{c}(\tau)\\
-i\sin(\varepsilon\mu^{2}\tau) & i\mathfrak{g}_{s}(\tau) & -\mathfrak{g}%
_{c}(\tau) & \cos(\varepsilon\mu^{2}\tau)
\end{pmatrix}
+\mathcal{O}\left(  \mu^{2}\right)  \text{,} \label{Theta4}%
\end{equation}
where we have defined the functions%
\begin{align*}
\mathfrak{g}_{c}(\tau)  &  =\mu\left[  \cos(\varepsilon\mu^{2}\tau
)-\mathfrak{h}_{c}(\tau)\right]  ,\\
\mathfrak{g}_{s}(\tau)  &  =\mu\left[  \sin(\varepsilon\mu^{2}\tau
)-\mathfrak{h}_{s}(\tau)\right]  ,\\
\mathfrak{h}_{c}(\tau)  &  =\exp\left[  -\left(  \eta+i\mu^{-1}\right)
\tau\right]  \cos\left(  \varepsilon\tau\right)  ,\\
\mathfrak{h}_{s}(\tau)  &  =\exp\left[  -\left(  \eta+i\mu^{-1}\right)
\tau\right]  \sin[\left(  \varepsilon\tau\right)  .
\end{align*}

From the condition for transfer of the state of the first oscillator to the
last one, fixed by the matrix structure (\ref{theta}), and assuming the
additional restriction $\left(  \varepsilon\mu\right)  ^{2}\ll1$, we thus
obtain from Eq. (\ref{Theta4}) the scaled exchange time
\[
\tau_{\mathrm{ex}}^{(N=4)}\simeq\frac{\pi}{2\varepsilon\mu^{2}}\left[
1+\mathcal{O}(\mu^{2})\right]  ,
\]
which implies the relations
\begin{subequations}
\label{Zeta}%
\begin{align}
\zeta_{1}^{r}\left(  \tau_{\mathrm{ex}}^{(N=4)}\right)   &  \simeq e^{-\eta
\mu^{2}\tau_{\mathrm{ex}}^{(4)}}\left[  \beta_{4}^{r}-\left(  i\mathfrak{g}%
_{c}(\tau_{\mathrm{ex}}^{(4)})\beta_{2}^{r}+\mathfrak{g}_{s}(\tau
_{\mathrm{ex}}^{(4)})\beta_{3}^{r}\right)  \right]  +\mathcal{O}\left(
\mu^{2}\right)  \text{,}\label{Zeta1}\\
\zeta_{4}^{r}\left(  \tau_{\mathrm{ex}}^{(N=4)}\right)   &  \simeq e^{-\eta
\mu^{2}\tau_{\mathrm{ex}}^{(4)}}\left[  \beta_{1}^{r}-\left(  \mathfrak{g}%
_{s}(\tau_{\mathrm{ex}}^{(4)})\beta_{2}^{r}+i\mathfrak{g}_{c}(\tau
_{\mathrm{ex}}^{(4)})\beta_{3}^{r}\right)  \right]  +\mathcal{O}\left(
\mu^{2}\right)  \text{,} \label{Zeta2}%
\end{align}
to ensure that the commutation $\left[  \Theta(\tau_{\mathrm{ex}}%
),\mathbf{H}^{D}\right]  =0$ is satisfied.

As is evident from the above expressions, the relaxation process represented
by the scaled damping rate $\eta$ prohibits a perfect state transfer,
attenuating the value of the excitations $\beta_{1}^{r}$ and $\beta_{4}^{r}$.
Moreover, the reservoirs also spoil the desired relation $\zeta_{1(4)}%
^{r}\left(  \tau_{\mathrm{ex}}^{(N=4)}\right)  =\beta_{4(1)}^{r}$ by mixing it
with the excitations of the nonideal transmitter oscillators. However, and
this is the core of our technique, the tunneling mechanism of state transfer
prompts the decay function $e^{-\eta\mu^{2}\tau_{\mathrm{ex}}^{(4)}}%
=e^{-\pi\Gamma/2\varepsilon\lambda}$ which approaches unity ---within the
ranges of the parameters outlined above, i.e., $\mu,\eta,\left(
\varepsilon\mu\right)  ^{2}\ll1$--- as the coupling $\varepsilon\lambda$
between the transmitting oscillators is increased. We note that the increase
of $\varepsilon\lambda$ decreases the exchange time\ $\tau_{\mathrm{ex}%
}^{(N=4)}\simeq\pi\Delta_{-}/2\mu\varepsilon\lambda$, and that is the reason
for the choice of a strong coupling strength $\varepsilon\lambda$ between the
transmitter oscillators: to enable the control of the exchange time
$\tau_{\mathrm{ex}}$, decreasing it as the fidelity of the transfer process
increases. Otherwise, without the dimensionless parameter $\varepsilon$,
increasing of the detuning $\Delta_{-}$ (i.e., the process allowing a
significant fidelity of the transferred state), would result in an
uncontrolled rise in the exchange time $\tau_{\mathrm{ex}}^{(N=4)}\simeq
\pi\Delta_{-}/2\mu\lambda$. From Eq. (\ref{Zeta}), we conclude that the
fidelity of the state transfer mechanism is maximized when the transmitter
oscillators are prepared in the vacuum state, apart from weakening the
inevitable system-reservoir coupling. In fact, the amount of excitation of the
dissipative transmission channel is directly proportional to the intensity of
noise injected into the transferred state. We finally note that for the case
of an ideal transmission channel, i.e., $\eta=0$, we obtain --- up to first
order corrections in $\mu$ carried out in $\mathfrak{g}_{c}(\tau_{\mathrm{ex}%
}^{(4)})$ and $\mathfrak{g}_{s}(\tau_{\mathrm{ex}}^{(4)})$ in the general case
of an excited transmission channel --- the desired relations $\zeta_{1}%
^{r}\left(  \tau_{\mathrm{ex}}^{(N=4)}\right)  \simeq\beta_{4}^{r}$ and
$\zeta_{4}^{r}\left(  \tau_{\mathrm{ex}}^{(N=4)}\right)  \simeq\beta_{1}^{r}$.

On the basis of the exchange time $\tau_{\mathrm{ex}}=\pi/2$ for the simplest
network, composed of two oscillators with coupling strength $\lambda$
\cite{Mickel2}, it is useful to assign an effective coupling strength to the
network which, for the case $N=4$, turns out to be $\lambda_{\mathrm{eff}%
}^{(N=4)}\approx\varepsilon\mu^{2}\lambda$.

\subsection{Nonideal linear network of $N=3,5,$ and $6$ oscillators}

Following the steps described above for the case $N=4$ and adopting the same
regime of parameters $\mu,\eta,\left(  \varepsilon\mu\right)  ^{2}\ll1$, we
obtain for $N=3$ the result%
\end{subequations}
\[
\tau_{\mathrm{ex}}^{(N=3)}\simeq\frac{\pi}{2\mu}\left[  1+\mathcal{O}(\mu
^{2})\right]  \text{,}%
\]
giving the expected effective coupling strength $\lambda_{\mathrm{eff}%
}^{(N=3)}\approx2\mu\lambda$ between the sender and receiver oscillators.
Evidently, it is also possible to derive analytical expressions for the scaled
exchange time for networks with $N>4$ whenever the diagonalization of the
associated matrix $\mathbf{H}_{lin(N)}^{D}$ generates a characteristic
polynomial that factorizes into parts of degree $\leq$ $4$. We have found that
$N=8$ is the limiting case permitting analytical solution, with a
characteristic polynomial that factorizes into two parts of degree $4$. For
$N=9$, the characteristic polynomial factorizes into two parts of degrees
$4\times5$. Analyzing the cases $N=5$ and $N=6$, which factorize into
polynomial of degrees $2\times3$ and $3\times3$, respectively, we obtain the
results%
\begin{align*}
\tau_{\mathrm{ex}}^{(N=5)}  &  \simeq\frac{\pi}{2\varepsilon^{2}\mu^{3}%
}\left[  1+\mathcal{O}(\mu^{2})\right]  ,\\
\tau_{\mathrm{ex}}^{(N=6)}  &  \simeq\frac{\pi}{2\varepsilon^{3}\mu^{4}%
}\left[  1+\mathcal{O}(\mu^{2})\right]  \text{.}%
\end{align*}
\qquad

We finally note that, without the imposition of the restriction $\left(
\varepsilon\mu\right)  ^{2}\ll1$, the above expressions for $\tau
_{\mathrm{ex}}$ would be approximately rewritten, for small values of $N$, as%
\begin{equation}
\tau_{\mathrm{ex}}^{(N)}\simeq\frac{\pi}{2\varepsilon^{(N-3)}\mu^{(N-2)}%
}\left\{  1+\left[  \mathcal{A}+\mathcal{B}\eta^{2}-\mathcal{C}\varepsilon
^{2}\right]  \mu^{2})\right\}  \text{,} \label{G}%
\end{equation}
where $\mathcal{A}=N-1$, $\mathcal{B}=%
{\textstyle\sum\nolimits_{m=1}^{N-2}}
m$, and $\mathcal{C=}N-3$. Evidently, for small $N$ and $\left(
\varepsilon\mu\right)  ^{2}\ll1$ the above expression is equivalent to the
results derived from $N=3$ to $N=6$. The derivation of an analytical
expression for $\tau_{\mathrm{ex}}^{(N)}$ in the general case of any $N$ is
not an easy task. As $N$ increases, we find that an involved dependence of the
second-order correction $\mathcal{O}(\mu^{2})$ on $N$ begins to play a
significant role. Although the task of identifying this dependence is still a
compelling challenge, in the present paper we analyze QPST for large values of
$N$ numerically, using expression (\ref{F}) to obtain the fidelity of the
transfer process.

\section{Numerical treatment of QPST in large nonideal linear networks}

\label{sec5}

Now, using the probability of a successful transfer of the desired state,
given by Eq. (\ref{F}), we analyze the QPST for large nonideal linear
networks. Consider the transfer of state $\mathcal{N}\left(  \left\vert
\alpha\right\rangle _{1}+\left\vert -\alpha\right\rangle _{1}\right)  $ to the
$N$th oscillator, with $\alpha=5$ and all other oscillators in the vacuum
state. In Fig. 1(a, b, c, and d) we plot the numerical curves for the exchange
probability $\mathcal{P}_{\mathrm{ex}}(\tau)$ against $\tau$ for the cases
$N=5,10,50,$ and $100$, respectively. Considering the above regime of
parameters, $\mu,\eta\ll1$ ---without requiring the additional restriction
$\left(  \varepsilon\mu\right)  ^{2}\ll1$--- we assume, in units of the
coupling strength $\lambda$, the fictitious value $\varpi=\omega=10$, giving
$\Delta_{-}=\mu^{-1}=10^{4}$ and $\varepsilon=5\times10^{3}$, apart from
$\eta=\Gamma=10^{-3}$. We first observe, as expected, that the exchange time
$\tau_{\mathrm{ex}}^{(N)}$ increases proportionally to $N$, being around
$\tau_{\mathrm{ex}}^{(N)}\approx\pi\times10^{4}$ for the case of $N=5$, in
agreement with the analytical result computed in Eq. (\ref{G}). Moreover, as
the state to be transferred occupies the virtual nonideal channel for a time
interval proportional to $N$, the fidelity of the transfer process decreases
with $N$, as displayed in Fig. 1. In fact, for the choice of decay rate
outlined above, we show that the fidelity of the transfer process is about
unity for the cases $N=5$ and $10$, beginning to exhibit a significant
decrease from $N=50$. We have not found any sensible decrease of the fidelity
for the cases $N=5$ and $10$, even for time intervals many orders of magnitude
longer than the exchange time. The shaded regions in the figures follow from
the strong oscillations of the probability $\mathcal{P}_{\mathrm{ex}}(\tau)$,
coming from the natural frequencies of the oscillators.

Focusing on the case $N=10$, in Fig. 2(a) we plot the exchange probability
$\mathcal{P}_{\mathrm{ex}}(\tau)$ against $\tau$, considering the same
parameters as in Fig. 1, except for the excitation $\beta=5$ of the coherent
states assumed to populate the oscillators composing the transmitter channel.
As expected, we verify that the exchange times are exactly those of the
associated case in Fig. 1(b). However, the pattern of the curves changes,
giving rise to the expected background oscillations due to the initial
excitation of the transmitter oscillators. Moreover, additional peaks occurs
at the recurrence time $\tau_{\mathrm{rec}}$, when the superposition state
$\mathcal{N}\left(  \left\vert \alpha\right\rangle _{1}+\left\vert
-\alpha\right\rangle _{1}\right)  $ goes back to the sender oscillator. The
magnitudes of these secondary peaks follow from the probability of finding the
superposition state in the receiver oscillator at the recurrence time, given
by $\mathcal{P}(\tau_{\mathrm{rec}})=\operatorname*{Tr}\left[  \rho_{1}%
(0)\rho_{N}(\tau_{\mathrm{rec}})\right]  =1/2$, where $\rho_{1}(0)$ and
$\rho_{N}(\tau_{\mathrm{rec}})$ represent the superposition states standing in
the first oscillator at $t=0$ and in the $N$th one at $t_{\mathrm{rec}}$. In
Fig. 2(b) we assume the same parameters as in Fig. 2(a) to plot the
probability of recurrence of the superposition state back to the first
oscillator, given by the expression $\mathcal{P}_{\mathrm{rec}}(\tau
)=\operatorname*{Tr}\left[  \rho_{1}(0)\rho_{1}(\tau)\right]  $. As expected,
in the recurrence time the superposition state recurs back to the sender oscillator.

In Fig. 3(a) we adopt the same parameters as in Fig. 1(b), except for the
smaller value $\varepsilon=8\times10^{2}$, to illustrate the expected increase
in the exchange time. An interesting feature of this figure is that the
fidelity of the transfer process does not depends on $\varepsilon$, despite
the increase in the exchange time, which is around 7 orders of magnitude
greater than in Fig. 1(b). These features are reinforced in Fig. 3(b) where,
again with the same parameters as in Fig. 1(b), we consider the limiting case
$\varepsilon=1$, which results in a formidable increase in the exchange time
of around 28 orders of magnitude, still preserving the fidelity.

With the same parameters as in Fig. 3(a), except for the smaller value of
$\Delta_{-}=2\times10^{3}$, in Fig. 4(a) we first see that the smaller
detuning pulls the exchange time back, to about the order of magnitude found
in Fig. 1(b), in spite of the quantity $\varepsilon=8\times10^{2}$. Moreover,
we verify the expected continuous decrease

of the process due to such a small detuning, which compels\ the state to
populate more effectively the virtual transmitter channel. In Fig. 4(b) we use
the same parameters as in Fig. 4(a), except for the coherent state $\alpha
=10$, to show that a larger excitation of the state to be transferred results
in a smaller fidelity of the process. In fact, the decoherence time of a
quantum state varies inversely with its excitation, in accordance with the
correspondence principle.

Finally, to illustrate the advantage of our tunneling-based scheme for state
transfer over those where the transfer proceeds non-virtually through all the
transmitter oscillators, in Fig. 5 we plot the exchange probability
$\mathcal{P}_{\mathrm{ex}}(\tau)$ against $\tau$ for the case where the sets
of parameters $\left\{  \omega_{m}\right\}  $ and $\left\{  \lambda
_{mn}\right\}  $ in Eq. (\ref{PP}) are utilized. Assuming, as in Fig. 1(a),
$\omega=10$ and $\eta=\Gamma=10^{-3}$, we find that, whereas the fidelity of
our tunneling scheme is about unity up to at least 10 times the exchange time
$\tau_{\mathrm{ex}}^{(N=5)}\approx7\times10^{5}$, that for the case of Fig. 5
decays to around zero for $\tau\approx2\times10^{3}$.

\section{Concluding remarks}

\label{sec6}

In this work we have discussed the problem of state transfer in a linear chain
of quantum dissipative harmonic oscillators. Assuming the first and last
oscillators to be on-resonant with each other and significantly off-resonant
with the transmitter channel ---running from the second to the last-but-one
oscillator--- we take advantage of the tunneling effect to circumvent the
decoherence during the transmission process. We have assumed ideal sender and
receiver oscillators connected by nonideal transmitters. Apart from the
significant improvement of the fidelity of our quasi-perfect transfer process
compared to those where the transfer proceeds through a non-virtual mechanism,
we have presented additional contributions to state transferring processes.

On the basis of a recently presented general treatment of a network of coupled
dissipative quantum harmonic oscillators, we have also derived a general
criterion for PST, whatever the topology of the network, which is applied here
to the particular case of a linear chain. Our criterion, given essentially by
the commutation relation between the network Hamiltonian $\mathbf{H}^{D}$ and
the matrix $\Theta(\tau_{\mathrm{ex}})$ connecting the initial states of the
network to those at the exchange time (when the transfer process is supposed
to be accomplished), has the advantage of being easier to handle
mathematically than that outlined in Ref. \cite{KostakNJ}. In fact, the matrix
$\mathbf{\Theta}(t)$, defined in Ref. \cite{MickelRedeGeral}, can be used for
two purposes: to ensure PST (or QPST) and to compute the exchange time
$\tau_{\mathrm{ex}}$, which can also be controlled by manipulating the
coupling between the transmitter oscillators. As a matter of fact, a large
detuning between the on-resonant and the off-resonant oscillators ensures not
only a high fidelity for the transfer process but also a significant
prolongation of the exchange time. In this connection, an increase in the
coupling between the transmitter oscillators acts to shorten such delays. The
role played by each of the network parameters is analyzed in details in a set
of figures presented after the formal development of our scheme.

We thus propose that the tunneling scheme presented here can be useful for
transferring quantum states between distant nodes of a quantum circuit without
their undergoing a significant coherence decay. Evidently, the control of the
network parameters, such as the natural frequencies and coupling strengths of
the oscillators, still represents a sensitive issue to be overcome
experimentally. We also point out that recent developments in circuit QED and
photonic crystals signal realistic platforms for the experimental
implementation of our proposed scheme.

\begin{flushleft}
{\Large \textbf{Acknowledgements}}
\end{flushleft}

The authors acknowledge the support from FAPESP and CNPQ, Brazilian agencies,
and thank F. L. Semi\~{a}o and G. D. de Moraes Neto for helpful discussions.

Fig. 1. Numerical plots of the exchange probability $\mathcal{P}_{\mathrm{ex}%
}(\tau)$ against $\tau$ for the transfer of the state $\mathcal{N}\left(
\left\vert \alpha\right\rangle _{1}+\left\vert -\alpha\right\rangle
_{1}\right)  $ to the $N$th oscillator, with $\alpha=5$ and all other
oscillators in the vacuum state. In units of the coupling strength $\lambda$
we set the fictitious value $\varpi=\omega=10$, prompting $\Delta_{-}=\mu
^{-1}=10^{4}$ and $\varepsilon=5\times10^{3}$, apart from $\eta=\Gamma
=10^{-3}$. Curves (a)-(d) refer to the cases $N=5,10,50,$ and $100$, respectively.

Fig. 2. (a) Numerical plots of the exchange probability $\mathcal{P}%
_{\mathrm{ex}}(\tau)$ against $\tau$ for the transfer of the state
$\mathcal{N}\left(  \left\vert \alpha\right\rangle _{1}+\left\vert
-\alpha\right\rangle _{1}\right)  $ to the $N$th oscillator, assuming the same
parameters as in Fig. 1, except for the excitation $\beta=5$ of the coherent
states populating the oscillators of the transmitter channel. (b) Probability
of recurrence $\mathcal{P}_{\mathrm{rec}}(\tau)$ of the initial superposition
back to the first oscillator, plotted against $\tau$.

Fig. 3. Numerical plots of the exchange probability $\mathcal{P}_{\mathrm{ex}%
}(\tau)$ against $\tau$ for the transfer of the state $\mathcal{N}\left(
\left\vert \alpha\right\rangle _{1}+\left\vert -\alpha\right\rangle
_{1}\right)  $ to the $N$th oscillator, assuming the same parameters as in
Fig. 1(b), except for the smaller values of the parameter (a) $\varepsilon
=8\times10^{2}$ and (b) $\varepsilon=1$.

Fig. 4. (a) Numerical plot of the exchange probability $\mathcal{P}%
_{\mathrm{ex}}(\tau)$ against $\tau$ for the transfer of the state
$\mathcal{N}\left(  \left\vert \alpha\right\rangle _{1}+\left\vert
-\alpha\right\rangle _{1}\right)  $ to the $N$th oscillator, assuming the same
parameters as in Fig. 3(a), except for the smaller value of the parameter
$\Delta_{-}=2\times10^{3}$. (b) The same as in (a), but with the coherent
state $\alpha=10$, to illustrate that the fidelity of the process decreases as
the excitation of the state to be transferred increases.

Fig. 5. Plot of the exchange probability $\mathcal{P}_{\mathrm{ex}}(\tau)$
against $\tau$ for the case where the network parameters in Eq. (\ref{PP}) are
assumed, together with $\omega=10$ and $\eta=\Gamma=10^{-3}$.

\end{document}